\documentstyle[12pt]{article}

\newlength{\titlesep}
\newlength{\authorsep}
\makeatletter

\renewcommand{\thesection}{\Roman{section}}

\def\fnum@figure{FIG.~\thefigure}
\newcommand{\Fig}[1]{\item[{\bf FIG.~\protect\ref{#1}}]}
\newcommand{\reffig}[1]{Fig.~\protect\ref{#1}}

\newcounter{figureparent}
\@addtoreset{figure}{figureparent}

\@addtoreset{equation}{section}
\renewcommand{\theequation}{(\arabic{section}.\arabic{equation})}
\def\@eqnnum{{\rm \theequation}}

\newcounter{eqnparent}
\@addtoreset{equation}{eqnparent}

\renewcommand{\abstract}{\if@twocolumn
  \section*{Abstract}
  \else
  \begin{center}
    {\bf Abstract\vspace{-.5em}\vspace{0pt}}
  \end{center}
  \fi}
\renewcommand{\endabstract}{\if@twocolumn\else\endquotation\fi}

\renewcommand{\appendix}{\par
  \setcounter{section}{0}
  \setcounter{subsection}{0}
  \renewcommand{\thesection}{Appendix~\Alph{section}}
  \renewcommand{\theequation}{(\Alph{section}.\arabic{equation})}}

\newcommand{\thismonth}{\ifcase\month\or
 January\or February\or March\or April\or May\or June\or
 July\or August\or September\or October\or November\or December\fi
 \space \number\year}

\newcommand{\preprintnumber}[1]
{\begin{flushright}
  \begin{tabular}{l} #1 \end{tabular}
  \end{flushright}}

\makeatother

\newcommand{\Rn}[1]{{\uppercase\expandafter{\romannumeral#1}}}
\newcommand{\gsim}%
{\mathrel{\mbox{\raisebox{-1.0ex}%
{$\stackrel{\textstyle >}{\textstyle \sim}$}}}}
\newcommand{\lsim}%
{\mathrel{\mbox{\raisebox{-1.0ex}%
{$\stackrel{\textstyle <}{\textstyle \sim}$}}}}
\newcommand{\ie}{{\it i.e.\/}\ }
\newcommand{\etc}{{\it etc.\/}\ }
\newcommand{\etal}{{\it et al.\/}\ }
\newcommand{\vev}[1]{\left\langle #1 \right\rangle}
\newcommand{\e}{{\rm e}}
\newcommand{\ol}[1]{\overline{#1}}
\newcommand{\wt}[1]{\widetilde{#1}}
\newcommand{\hc}{{\rm h.\ c.}\ }
\newcommand{\bsg}{$b\rightarrow s\ \gamma$\ }

\newcommand{\bb}{$B^0$--$\ol{B}^0$\ }
\newcommand{\kk}{$K^0$--$\ol{K}^0$\ }

\newcommand{\Journal}[4]{{#1} {\bf #2}, {#4} {(#3)}}
\newcommand{\pl}{\sl Phys.~Lett.}
\newcommand{\plb}{\sl Phys.~Lett.~{\bf B}}
\newcommand{\prp}{\sl Phys.~Rep.}

\newcommand{\prd}{\sl Phys.~Rev.~{\bf D}}
\newcommand{\prl}{\sl Phys.~Rev.~Lett.}
\newcommand{\np}{\sl Nucl.~Phys.}
\newcommand{\npb}{\sl Nucl.~Phys.~{\bf B}}
\newcommand{\ptp}{\sl Prog.~Theor.~Phys.}

\newcommand{\zpc}{\sl Z.~Phys.~{\bf C}}
\newcommand{\mpl}{\sl Mod.~Phys.~Lett.}
\newcommand{\mpla}{\sl Mod.~Phys.~Lett.~{\bf A}}

\newcommand{\ibid}{\it ibid.}

\newcommand{\epsfile}[1]{\relax}

\begin{document}
\baselineskip 18pt

\begin{titlepage}
\preprintnumber{%
KEK-TH-445 \\
KEK preprint 95-125 \\
ICRR-Report-341-95-7 \\
TU-485 \\
hep-ph/9510286 \\
October 1995
}
\vspace*{\titlesep}
\begin{center}
{\LARGE\bf
\bb mixing and $\epsilon_K$ parameter in the
minimal supergravity model
}
\\
\vspace*{\titlesep}
{\large Toru {\sc Goto}$^1$},
{\large Takeshi {\sc Nihei}$^2$}
and
{\large Yasuhiro {\sc Okada}$^3$}\\
\vspace*{\authorsep}
{\it $^1$Department of Physics, Tohoku University \\
  Sendai 980-77 Japan}
\\
\vspace*{\authorsep}
{\it $^2$Institute for Cosmic Ray Research, University of Tokyo \\
  3-2-1 Midori-cho, Tanashi-shi, Tokyo 188 Japan}
\\
\vspace*{\authorsep}
{\it $^3$Theory Group, KEK, Tsukuba, Ibaraki, 305 Japan }
\end{center}
\vspace*{\titlesep}
\begin{abstract}
\bb mixing and a CP violation parameter in \kk mixing $\epsilon_K$ are
studied in the minimal supergravity model.
We solve the one-loop renormalization group equations for the minimal
supersymmetric standard model (MSSM) parameters numerically
in order to determine the masses and mixings of the supersymmetric
particles, while all off-diagonal (generation mixing) elements and
phases of Yukawa coupling matrices and those of squark mass matrices are
taken into account.
Applying the radiative electroweak symmetry breaking condition and
phenomenological constraints including the recent measurement of the
\bsg inclusive branching ratio, we obtain the allowed parameter region.
We have found that the present constraints still allow a parameter
region where both \bb mass splitting $\Delta M_B$ and $\epsilon_K$ are
$\sim 20$\% larger than the standard model values.
By explicit numerical calculations, we have also found that
the complex phase of \bb mixing matrix element in this model is almost
the same as the standard model value in a good accuracy in the whole
allowed parameter region.
It is shown that $\Delta M_B$ and $\epsilon_K$ can put useful
constraints on the supersymmetry's parameter space when the
Cabibbo-Kobayashi-Maskawa matrix elements are determined through the
measurements of CP violations in $B$ decay with future $B$-factories.
\end{abstract}

\end{titlepage}

%%%%%%%%%%%%%%%%%%%%%%%%%%%%%%%%%%%%%%%%%%%%%%%%%%%%%%%%%%%%%%%%%%%%%%

\section{Introduction}
\label{sec:intro}

Supersymmetry (SUSY) is one of the most favorable candidate for new
physics beyond the standard model (SM).
The minimal supersymmetric standard model (MSSM), which is the most
straightforward supersymmetric extension of SM, has been intensively
studied for years.
MSSM has many new particles such as superpartners of ordinary
particles and extra Higgs bosons.
Though these SUSY particles are sufficiently heavy to evade the direct
search in the present accelerator experiments, they may give measurable
contributions to low energy phenomena such as flavor changing neutral
currents (FCNC) and CP violations through the radiative corrections
\cite{kkbar,edm,bks,bbmr,bsgamma,go,bbbar,bcko}.

MSSM has two kinds of sources of flavor mixings: a Yukawa coupling
sector and a SUSY breaking sector.
The Higgs sector of MSSM is a special case of two Higgs doublet model
(THDM) categorized as model II \cite{thdm2}, in which up-type (electric
charge $2/3$) quarks and down-type (electric charge $-1/3$) quarks get
masses from vacuum expectation values of different Higgs doublet fields
and hence there is no tree level FCNC.
After the diagonalization of quark mass matrices, flavor mixing between
quarks appears in the coupling with $W$ boson, just as in the SM, and in
the coupling with the physical charged Higgs boson.
Both flavor mixings are described by the Cabibbo-Kobayashi-Maskawa (CKM)
matrix with three mixing angles and one CP violating complex phase.

The other source of flavor mixing and CP violation lies in the SUSY
breaking sector which involves SUSY particles.
Since SUSY is softly broken, squark masses do not have to be
diagonalized in the same flavor basis as that for quarks.
For a general SUSY breaking sector, the SUSY contributions to the FCNC
and/or CP violating processes can easily dominate over
the SM contributions.
Such a model is strongly constrained by the present
experiments on the \kk mixing \cite{kkbar} and the neutron electric
dipole moment \cite{edm}.

Minimal supergravity (SUGRA) provides an attractive framework for the
SUSY breaking sector of MSSM \cite{sugra}.
In the minimal SUGRA model, SUSY is assumed to be spontaneously broken
in the ``hidden'' sector which couples to the ``observable'' MSSM sector
only gravitationally so that the interactions between the hidden and the
observable sectors are suppressed by $O(M_{\rm Planck}^{-1})$.
The induced soft SUSY breaking terms have a universal structure: all
soft SUSY breaking masses of squarks and sleptons are degenerate, all
trilinear scalar couplings are proportional to the corresponding Yukawa
couplings, \etc
In such a case, the flavor mixings in both quark sector and squark
sector are essentially determined by the CKM parameters.
Imposing the universal structure on the soft SUSY breaking parameters at
a high energy scale such as the GUT scale, one can evaluate the soft
SUSY breaking parameters and the Higgs potential at the electroweak
scale by solving renormalization group equations (RGEs).
As a result of the renormalization effects by the large third generation
Yukawa couplings, the Higgs potential is modified so that the
electroweak symmetry breaking  occurs.
This is called the radiative electroweak symmetry breaking scenario
\cite{radbr}.
Flavor mixings in the squark sector are also determined by the RGEs.
The first and the second generations of squarks
are highly degenerate so that the constraint from the $K_L$--$K_S$ mass
splitting is easily satisfied.
Furthermore, if we assume that all SUSY breaking parameters are real at
the GUT scale, the neutron electric dipole moment is sufficiently
suppressed \cite{edm}.

Although many analyses of the minimal SUGRA model based on the above
scenario have been published in literature
\cite{kkbar,edm,bks,bbmr,bsgamma,go,bbbar,bcko,sugra}, an extensive
study on the FCNC processes is missing which takes into accounts the
radiative electroweak symmetry breaking scenario and recent experimental
results, such as the determination of the top quark mass \cite{top} and
the measurement of \bsg inclusive branching ratio \cite{cleo}.
In particular, from the recent theoretical studies on the \bsg process
\cite{bsgamma}, it is shown that relatively light charged Higgs and/or
SUSY particles are still allowed since the SUSY particles' contributions
to this process can cancel the charged Higgs contribution depending on
the sign of the higgsino mass parameter.
Therefore, it is important to determine how such light charged Higgs
and/or SUSY particles contribute to other FCNC processes.

The purpose of the present paper is to study the FCNC and CP violation
in the framework of the minimal SUGRA model.
We focus on three quantities:
the complex phase of the \bb mixing matrix element $M_{12}(B) \equiv
\e^{i\theta_B} |M_{12}(B)|$, which is related to the CP violation in $B$
meson decays,
\bb mass splitting $\Delta M_B = 2|M_{12}(B)|$, and the $\epsilon_K$
parameter of the CP violation in the \kk system.
Unlike the $K_L$--$K_S$ mass splitting $\Delta M_K$ in which the long
distance contribution cannot be neglected, $\Delta M_B$ and $\epsilon_K$
are supposed to be dominated by the short distance physics
\cite{kkbarWolf}, thus can be sensitive to new physics contributions.
In most of the previous works \cite{bks,bbbar,bcko}, \bb and \kk mixing
in MSSM are studied with some simplified treatments on the SUSY particle
masses and mixing angles, such as:
approximate solutions of the RGEs which are obtained by neglecting
Yukawa couplings other than that of top quark are used; or
a simple form of mass matrices at the electroweak scale is assumed.
On the contrary, in the present analysis, we obtain all mass matrices of
SUSY particles from the universal soft SUSY breaking parameters at the
GUT scale by a straightforward numerical calculation.
We include all complex elements of Yukawa coupling matrices and of
squark mass matrices in solving one-loop RGEs for all MSSM parameters
with the universal boundary conditions explained above.
We then evaluate the effective potential for the Higgs fields at the
electroweak scale to find a consistent SU(2)$\times$U(1) breaking
minimum in accordance with the radiative electroweak symmetry breaking
scenario.
The obtained mass matrices of all particles are diagonalized to evaluate
the flavor mixing in the squark sector.
The mass spectrum and the mixing are then used to calculate \bb and \kk
mixing matrix elements.
Along this outline, FCNC processes in $B$ decays and \bb mixing in the
minimal SUGRA model are studied previously in Ref.~\cite{bbmr}.
Compared with this work, we improve the following points:
$\epsilon_K$ is also considered;
one-loop correction to the effective Higgs potential \cite{oyy} is
included to determine the electroweak symmetry breaking;
no special relation between soft SUSY breaking parameters $A$ and $B$
(see Sec.~\ref{sec:model}) is assumed;
and experimental constraints by LEP \etc are updated \cite{CDF,LEP}, as
well as the top quark mass and \bsg branching ratio.
It is found that the SUSY contributions do not change the phase of the
\bb mixing $\theta_B$ from the SM value appreciably
for the whole SUSY breaking and CKM parameter space we considered.
As for $\Delta M_B$ and $\epsilon_K$, we find that
all the contributions from charged Higgs and SUSY particles have the
same sign as the SM contribution and that a parameter region in which
both $\Delta M_B$ and $\epsilon_K$ are $\sim 20$\% larger than the
standard model values is allowed by the present constraints.
We also find that there is a linear correlation between the ratio
of $\Delta M_B$ to its SM value and that of $\epsilon_K$.

The rest of the paper is organized as follows.
In the next section the minimal SUGRA model is introduced to clarify the
notations and the assumptions which we adopt in this paper.
In Sec.~\ref{sec:results}, our results of numerical analyses are
presented.
Sec.~\ref{sec:conclusion} is devoted for discussion and conclusions.
Formulae for functions from loop integrals and QCD correction factors are
summarized in the Appendices.

\section{Minimal SUGRA model}
\label{sec:model}

MSSM contains three generations of matter (left-handed) chiral
superfields
$Q_i$ ($   {\bf 3}$, ${\bf 2}$, $ 1/6$),
$D_i$ ($\ol{\bf 3}$, ${\bf 1}$, $ 1/3$)
and
$U_i$ ($\ol{\bf 3}$, ${\bf 2}$, $-2/3$)
for quark supermultiplets,
$L_i$ ($   {\bf 1}$, ${\bf 2}$, $-1/2$)
and
$E_i$ ($   {\bf 1}$, ${\bf 1}$, $ 1  $)
for lepton supermultiplets,
where SU(3)$\times$SU(2)$\times$U(1) quantum numbers are
expressed in each bracket and the suffix $i=1,2,3$ is the
generation index,
and two Higgs doublets
$H_1$ ($   {\bf 1}$, ${\bf 2}$, $-1/2$)
and
$H_2$ ($   {\bf 1}$, ${\bf 2}$, $ 1/2$),
as well as vector superfields for gauge multiplets.
Yukawa coupling and supersymmetric Higgs mass terms are
described by the superpotential $W_{\rm MSSM}$ as
\begin{equation}
    W_{\rm MSSM} = f_D^{ij} Q_i^{a\alpha} D_{ja} H_{1\alpha}
      + f_U^{ij} \epsilon_{\alpha\beta} Q_i^{a\alpha} U_{ja} H_2^\beta
      + f_L^{ij} \epsilon^{\alpha\beta} E_i L_{j\alpha} H_{1\beta}
      + \mu H_{1\alpha} H_2^\alpha ~,
  \label{eq:superpotential}
\end{equation}
where
$f_D$, $f_U$ and $f_L$ are Yukawa coupling constants for
down-type quarks, up-type quarks and leptons, respectively,
the suffices $a,b,c=1,2,3$ and $\alpha,\beta=1,2$ are SU(3)
are SU(2) indices, respectively.
$\epsilon_{\alpha\beta}$ and $\epsilon^{\alpha\beta}$ are antisymmetric
tensors with $\epsilon_{12} = \epsilon^{12} = 1$.

Throughout the calculation hereafter, we choose the basis in the
generation space for the superfields such that the Yukawa coupling
constants for up-type quarks $f_U$ and leptons  $f_L$  are to be
diagonal at the electroweak scale.
The Yukawa terms in \ref{eq:superpotential} are then written as
\begin{equation}
  W_{\rm Yukawa}(m_Z) =
         \hat{f}_D^{kj}
         \left( V_{\rm KM}^\dagger \right)_k^{~i} Q_i D_j H_1
       + \hat{f}_U^{ij} Q_i U_j H_2
       + \hat{f}_L^{ij} E_i L_j H_1 ~,
  \label{eq:YukawaZ}
\end{equation}
where the notation ``$\hat{~~}$'' stands for a diagonal matrix.
All eigenvalues of $\hat{f}_D$, $\hat{f}_L$ and $\hat{f}_U$ are taken
to be real positive.
We use the standard parameterization in Ref.~\cite{PDG}%
\footnote{
In the Wolfenstein parametrization \cite{Wolf} the $V_{\rm KM}$ is
parametrized by four parameters $(\lambda,~ A,~ \rho,~ \eta)$.
The parameters $\rho$ and $\eta$ are written as
$\rho + i \eta = -(V_{ub}^* V_{ud})/(V_{cb}^* V_{cd})$
neglecting the higher order terms of the Cabibbo angle
$\lambda = V_{us}$.
}
for the CKM matrix $V_{\rm KM}$.

In addition to the supersymmetric Lagrangian to be derived from
\ref{eq:superpotential}, the following soft SUSY breaking terms are
included:
\begin{eqnarray}
  -{\cal L}_{\rm soft}
  &=& (m_Q^2)^i_{~j} \wt{q}_i \wt{q}^{\dagger j}
    + (m_D^2)_i^{~j} \wt{d}^{\dagger i} \wt{d}_j
    + (m_U^2)_i^{~j} \wt{u}^{\dagger i} \wt{u}_j
  \nonumber\\
  &&+ (m_E^2)^i_{~j} \wt{e}_i \wt{e}^{\dagger j}
    + (m_L^2)_i^{~j} \wt{l}^{\dagger i} \wt{l}_j
  \nonumber\\
  &&+ \Delta_1^2 h_1^{\dagger} h_1
    + \Delta_2^2 h_2^\dagger h_2
    - \left( B\mu h_1 h_2 + \hc \right)
  \nonumber\\
  &&+ \left(   A_D^{ij} \wt{q}_i \wt{d}_j h_1
             + A_U^{ij} \wt{q}_i \wt{u}_j h_2
             + A_L^{ij} \wt{e}_i \wt{l}_j h_1
             + \hc \right)
  \nonumber\\
  &&+ \left(   \frac{M_1}{2} \wt{B}\wt{B}
             + \frac{M_2}{2} \wt{W}\wt{W}
             + \frac{M_3}{2} \wt{G}\wt{G} + \hc \right) ~.
  \label{eq:softbreaking}
\end{eqnarray}
where $\wt{q}_i$, $\wt{d}_i$, $\wt{u}_i$, $\wt{e}_i$,
$\wt{l}_i$, $h_1$ and $h_2$ are scalar components of $Q_i$, $D_i$,
$U_i$, $E_i$, $L_i$, $H_1$ and $H_2$, respectively, and $\wt{B}$,
$\wt{W}$ and $\wt{G}$ are U(1), SU(2) and SU(3) gauge fermion
fields (bino, wino and gluino), respectively.
SU(2) and SU(3) suffices are omitted in \ref{eq:softbreaking} for
simplicity.
In the minimal SUGRA model, SUSY is assumed
to be spontaneously broken in the hidden sector which
couples to the observable sector (MSSM in the present case)
only gravitationally, and hence universal soft SUSY breaking
terms are induced in the observable sector.
Here, we assume that the soft SUSY breaking parameters
satisfy the following relations at the GUT scale:
\begin{eqnarray}
  (m_Q^2)^i_{~j} &=&
  (m_E^2)^i_{~j} ~=~ m_0^2\ \delta^i_{~j} ~,
\nonumber\\
  (m_D^2)_i^{~j} &=&
  (m_U^2)_i^{~j} ~=~
  (m_L^2)_i^{~j} ~=~ m_0^2\ \delta_i^{~j} ~,
\nonumber\\
  \Delta_1^2 &=& \Delta_2^2 ~=~ m_0^2 ~,
\nonumber\\
  A_D^{ij} &=& f_{DX}^{ij} A_X ~, ~~
  A_L^{ij} ~=~ f_{LX}^{ij} A_X ~, ~~
  A_U^{ij} ~=~ f_{UX}^{ij} A_X ~,
  \nonumber\\
  M_1 &=& M_2 ~=~ M_3 ~=~ M_{gX} ~,
\label{eq:boundaryconditions}
\end{eqnarray}
where the suffix ``$X$'' stands for the value at the GUT scale.
We also assume that $A_X$, $M_{gX}$ and $\mu$ are all real parameters.
Therefore, no new CP violating complex phase (other than that in CKM
matrix) is introduced in the present analysis.
Although two physical complex phases among these soft SUSY breaking
parameters are possible in principle, such phases lead to a large
neutron electric dipole moment in general and are strongly constrained
\cite{edm}.

Below the GUT scale, radiative corrections modify all parameters in
the superpotential \ref{eq:superpotential} and the soft SUSY breaking
terms \ref{eq:softbreaking}, as well as three gauge coupling constants
$g_1$, $g_2$ and $g_3$ for U(1), SU(2) and SU(3), respectively.
The evolution of the parameters are described by the RGEs \cite{bks,bbmr}.
According to the radiative SU(2) $\times$ U(1) breaking scenario
\cite{radbr}, we numerically solve the RGEs down to the electroweak scale
$m_Z$ and evaluate the effective potential for the neutral Higgs
fields:
\begin{eqnarray}
  V({\rm Higgs}) &=& V_{\rm tree} + V_{\mbox{\scriptsize 1-loop}} ~,
  \nonumber\\
  V_{\rm tree} &=&   \left( \mu^2 + \Delta_1^2 \right) |h_1|^2
          + \left( \mu^2 + \Delta_2^2 \right) |h_2|^2
          - \left( B\mu h_1 h_2 + {\rm h.~c.} \right)
  \nonumber\\
    & & + \frac{g_1^2 + g_2^2}{8} \left( |h_1|^2 - |h_2|^2 \right)^2 ~,
  \label{Higgspotential}
\end{eqnarray}
where $V_{\mbox{\scriptsize 1-loop}}$ is the one-loop correction to the
effective potential induced by the Yukawa couplings for the third
generation \cite{oyy}.
We have imposed that the electroweak symmetry is broken properly and
gives the relation
\begin{eqnarray}
  \vev{h_1} &=& v \cos \beta ~, ~~
  \vev{h_2} ~=~ v \sin \beta ~,
  \label{vev}
  \\
  m_Z^2 &=& \frac{g_2^2}{2\cos^2\theta_W} v^2 ~,
  \nonumber
\end{eqnarray}
where $\theta_W$ is the Weinberg angle and $\beta$ is the angle for the
vacuum expectation values of the two Higgs fields.
The magnitudes of $\mu$ and $B$ in \ref{Higgspotential} are determined
by the condition \ref{vev}.

New flavor mixings in the quark--squark--gaugino and
quark--squark--higgsino couplings come from diagonalization of the
quark mass matrices as well as the squark ones.
The mass matrix for up-type squarks is expressed as
\begin{eqnarray}
  -{\cal L}(\mbox{s-up mass}) &=&
       ( \wt{q}_{u} ,~ \wt{u}^{\dagger} )
       {\cal M}_{\wt{u}}^2
        \left( \begin{array}{c}
                 \wt{q}_{u}^{\dagger} \\ \wt{u}
               \end{array}\right) ~,
  \nonumber\\
  &=&  ( \wt{q}_{ui} ,~ \wt{u}^{\dagger i} )
       \left( \begin{array}{cc}
                \left( m_{LL}^2 \right)^i_{~j} &
                \left( m_{LR}^2 \right)^{ij} \\
                \left( m_{RL}^2 \right)_{ij} &
                \left( m_{RR}^2 \right)_i^{~j}
              \end{array}\right)
              \left( \begin{array}{c}
                       \wt{q}_{u}^{\dagger j} \\ \wt{u}_j
                     \end{array}\right) ~,
  \nonumber\\
  \left( m_{LL}^2 \right)^i_{~j} &=&  \left( M_U M_U^\dagger \right)^i_{~j}
    + \left( m_Q^2 \right)^i_{~j}
    + m_W^2 \cos 2\beta \left(   \frac{1}{2}
                               - \frac{1}{6}\tan^2 \theta_W \right)
      \delta^i_{~j} ~,
  \nonumber\\
  \left( m_{RR}^2 \right)_i^{~j} &=& \left( M_U^\dagger M_U \right)_i^{~j}
    + \left( m_U^2 \right)_i^{~j}
    + m_W^2 \cos 2\beta \left( \frac{2}{3}\tan^2 \theta_W \right)
      \delta_i^{~j}   ~,
  \nonumber\\
  \left( m_{LR}^2 \right)^{ij} &=& \mu M_U^{ij} \cot\beta
    + A_U^{ij} v \sin\beta ~,
  \nonumber \\
  m^2_{RL} &=& m_{LR}^{2\dagger} ~,
  \label{squarkmass}
\end{eqnarray}
where $M_U$ is the up-type quark mass matrix $M_U^{ij} = f_U^{ij} v
\sin\beta$ and $\wt{q}_u$ is the up-type component of the SU(2)
doublet $\wt{q}$.
Note that even if we take the basis in which $M_U$ is diagonalized as
Eq.~\ref{eq:YukawaZ}, the squark mass matrix ${\cal M}_{\wt{u}}^2$ is
not diagonalized simultaneously since, due to the renormalization
effect, off-diagonal elements are induced in the soft SUSY breaking
parameter matrices.
Squark mass basis is obtained by diagonalizing \ref{squarkmass} with
a 6$\times$6 unitary matrix $\wt{U}_U$:
\begin{eqnarray}
  \wt{u}'_I &=& \left( \wt{U}_U \right)_I^{~J} \wt{u}_J ~,
  ~~ I ~=~ 1,\ 2,\ \cdots,\ 6 ~,
  \nonumber\\
  \wt{u}_I &=&
    \left\{ \begin{array}{lcl}
               \wt{q}_{uI} & \mbox{for} & I = 1,\ 2,\ 3 \\
               \wt{u}^\dagger_{I-3} & \mbox{for} & I = 4,\ 5,\ 6
            \end{array} \right. ~,
  \nonumber\\
  \wt{U}_U^\dagger {\cal M}_{\wt{u}}^{2{\bf T}} \wt{U}_U &=&
     \mbox{diagonal} ~,
  \label{DiagonalizeSquark}
\end{eqnarray}
where $\wt{u}'_I$ is the mass eigenstate of up-type squark and
$^{\bf T}$ stands for transposition.
The mass bases of down-type squarks are obtained
in the same way with 6$\times$6 unitary matrices $\wt{U}_D$.
The flavor mixings in the quark--squark--gaugino and the
quark--squark--higgsino coupling are described by the mixing matrices
$\wt{U}_U$, $\wt{U}_D$ and the CKM matrix.

\bb and \kk mixing matrix elements $M_{12}(B)$ and $M_{12}(K)$ are
evaluated with use of the box diagrams which contain various particles
in the internal loop.
In addition to the standard model contribution ($W$ and up-type quark
loops), the following diagrams contribute to $M_{12}(B/K)$ in MSSM:
\begin{enumerate}
\item charged Higgs -- up-type quark loops
  and charged Higgs -- $W$ -- up-type quark loops,
\item chargino -- up-type squark loops,
\item gluino -- down-type squark loops,
\item neutralino -- down-type squark loops
  and neutralino -- gluino -- down-type squark loops.
\end{enumerate}
The contribution from the box diagrams involving neutralinos is
estimated to be smaller than the gluino contribution and is neglected
in the present calculation.
Furthermore we neglect the contribution from diagrams with right-handed
external quark lines since the flavor mixing in the right-handed sector
is small in the minimal SUGRA model \cite{gna}.
Consequently, our formulae for $M_{12}(B)$ are given by the following
expressions.
In the standard model and the charged Higgs contributions
$A_{\rm SM}(B)$ and $A_{H^\pm}(B)$ we have taken the
$m_{u,c} \rightarrow 0$ limit since these contributions are negligible
compared to the top mass contributions:
\begin{subequations}
\begin{eqnarray}
M_{12}(B) &=&
  \frac{\hat{B}_B \eta_B f_B^2 M_B}{384\pi^2}
  \left[   A_{\rm SM}(B)   + A_{H^\pm}(B)
         + A_{\chi^\pm}(B) + A_{\wt{g}}(B) \right] ~,
\label{eq:bbtotal}
\\
A_{\rm SM}(B) &=&
  \frac{g_2^4}{m_W^2}
  \left( V_{td}^* V_{tb} \right)^2 F_1( x_t ) ~,
\label{eq:bbsm}
\\
A_{H^\pm}(B) &=&
  \frac{g_2^4}{m_W^2 \tan^2\beta}
  \left( V_{td}^* V_{tb} \right)^2 x_t^2
  \left[
    \frac{1}{4 x_H \tan^2\beta} G_1( x_t^H, x_t^H )
  \right.
\nonumber\\&&
  \left.
  + \frac{1}{2} G'_1( x_t, x_t, x_H )
  - 2           G'_0( x_t, x_t, x_H )
  \right] ~,
\label{eq:bbhc}
\\
A_{\chi^\pm}(B) &=&
  \sum_{\alpha,\beta=1}^{2} \sum_{I,J=1}^{6}
  \frac{g_2^2}{M_C^{\alpha 2}}
  \left[   \left( \wt{U}'^\dagger_U  \right)_1^{~I}
           \left( U_+^\dagger        \right)_{\alpha}^{~1}
         + \left( \wt{U}''^\dagger_U \right)_1^{~I}
           \left( U_+^\dagger        \right)_{\alpha}^{~2} \right]
\nonumber\\&& \times
  \left[   \left( \wt{U}'_U          \right)_J^{~3}
           \left( U_+                \right)_1^{~\alpha}
         + \left( \wt{U}''_U         \right)_J^{~3}
           \left( U_+                \right)_2^{~\alpha} \right]
\nonumber\\&& \times
  \left[   \left( \wt{U}'^\dagger_U  \right)_1^{~J}
           \left( U_+^\dagger        \right)_{\beta }^{~1}
         + \left( \wt{U}''^\dagger_U \right)_1^{~J}
           \left( U_+^\dagger        \right)_{\beta }^{~2} \right]
\nonumber\\&& \times
  \left[   \left( \wt{U}'_U          \right)_I^{~3}
           \left( U_+                \right)_1^{~\beta }
         + \left( \wt{U}''_U         \right)_I^{~3}
           \left( U_+                \right)_2^{~\beta } \right]
  G'_1( x_I^\alpha, x_J^\alpha, x_\beta^\alpha ) ~,
\label{eq:bbcno}
\\
A_{\wt{g}}(B) &=&
  \sum_{I,J=1}^{6} \frac{g_3^4}{M_3^2}
  \left( \wt{U}'^\dagger_D \right)_1^{~I}
  \left( \wt{U}'_D \right)_I^{~3}
  \left( \wt{U}'^\dagger_D \right)_1^{~J}
  \left( \wt{U}'_D \right)_J^{~3}
\nonumber\\&& \times
  \left\{
      \frac{22}{9} G_1( x_I^{\wt{g}}, x_J^{\wt{g}} )
    + \frac{ 8}{9} G_0( x_I^{\wt{g}}, x_J^{\wt{g}} )
  \right\} ~.
\label{eq:bbgl}
\end{eqnarray}
\end{subequations}
Here, the mixing matrices $\wt{U}'_{U,D}$ and $\wt{U}''_{U}$ are defined
as
\begin{subequations}
\begin{eqnarray}
  \left( \wt{U}'_{U,D} \right)_I^{~j} &\equiv&
  \left( \wt{U}_{U,D} \right)_I^{~k} \left( V_{\rm KM} \right)_k^{~j} ~,
\\
  \left( \wt{U}''_{U} \right)_I^{~j} &\equiv&
  \left( \wt{U}_{U} \right)_I^{~k+3}
  \frac{m^{(u)}_k}{\sqrt{2}m_W \sin\beta}
  \left( V_{\rm KM} \right)_k^{~j} ~,
\end{eqnarray}
\end{subequations}
where $m^{(u)}_k$ ($k=1,2,3$) is up-type quark mass;
$M_C^\alpha$ and $U_+$ are the eigenvalue and the diagonalizing matrix
of the chargino mass matrix ${\cal M}_C$:
\begin{eqnarray}
  U_-^\dagger {\cal M}_C U_+ &=&
 -\left(
    \begin{array}{cc}
      M_C^1 & 0 \\ 0 & M_C^2
    \end{array}
  \right) ~,
\nonumber\\
  {\cal M}_C &=&
  \left(
    \begin{array}{cc}
      M_2 & \sqrt{2}m_W \sin\beta \\
      -\sqrt{2}m_W \cos\beta & -\mu
    \end{array}
  \right) ~.
\end{eqnarray}
The variables $x$'s are defined as
$x_t = m_t^2 / m_W^2$,
$x_H = m_{H^\pm}^2 / m_W^2$,
$x_t^H = x_t/x_H$,
$x_I^\alpha = m_{\wt{u}'_I}^2 / M_C^{\alpha 2}$,
$x_\beta^\alpha = M_C^{\beta 2} / M_C^{\alpha 2}$ and
$x_I^{\wt{g}} = m_{\wt{d}'_I}^2/M_3^2$ with the gluino mass $M_3$,
and the Inami-Lim functions $F_{1,2}$, $G_{0,1}$
and $G'_{0,1}$ are listed in \ref{sec:ilf}.
We evaluate all masses and mixing matrices at the electroweak scale
neglecting the electroweak and SUSY threshold corrections.
Overall factors $\hat{B}_B$, $f_B$, $M_B$ and $\eta_B$ are the
bag parameter, decay constant, $B$ meson mass and the QCD factor below
the weak scale, respectively.
We use the one-loop formula for $\eta_B$ (see \ref{sec:qcdfactors}),
which is sufficient for the present purpose, since our main interest is
to study the ratio to the SM value and hence the overall factor is
irrelevant.
For $M_{12}(K)$, terms with the first order of $x_c = m_c^2(m_W)/m_W^2$
have to be included in the standard model contribution $A_{\rm SM}$ and
in the charged Higgs contribution $A_{H^\pm}$:
\begin{subequations}
\begin{eqnarray}
M_{12}(K) &=&
  \frac{\hat{B}_K \eta_K f_K^2 M_K}{384\pi^2}
  \left[   A_{\rm SM}(K)   + A_{H^\pm}(K)
         + A_{\chi^\pm}(K) + A_{\wt{g}}(K) \right] ~,
\label{eq:kktotal}
\\
A_{\rm SM}(K) &=&
  \frac{g_2^4}{m_W^2}
  \left\{
    \left( V_{cd}^* V_{cs} \right)^2 \hat{\eta}_1 x_c
  + \left( V_{td}^* V_{ts} \right)^2 F_1( x_t )
  \right.
\nonumber\\&&
  \left.
  +2\left( V_{cd}^* V_{cs} \right)
    \left( V_{td}^* V_{ts} \right)
    \left[ x_c F_2( x_t ) - \hat{\eta}_2 x_c \log x_c \right]
  \right\} ~,
\label{eq:kksm}
\\
A_{H^\pm}(K) &=&
  \frac{g_2^4}{m_W^2 \tan^2\beta}
  \left\{
    \left( V_{td}^* V_{ts} \right)^2 x_t^2
    \left[
        \frac{1}{4 x_H \tan^2\beta} G_1( x_t^H, x_t^H )
    \right.
  \right.
\nonumber\\&&
    \left.
      + \frac{1}{2} G'_1( x_t, x_t, x_H )
      - 2           G'_0( x_t, x_t, x_H )
    \right]
\nonumber\\&&
  +2\left( V_{cd}^* V_{cs} \right)
    \left( V_{td}^* V_{ts} \right) x_c x_t
    \left[
        \frac{1}{4 x_H \tan^2\beta} G_1( x_c^H, x_t^H )
    \right.
\nonumber\\&&
  \left.
    \left.
      + \frac{1}{2} G'_1( x_c, x_t, x_H )
      - 2           G'_0( x_c, x_t, x_H )
    \right]
  \right\} ~.
\label{eq:kkhc}
\end{eqnarray}
\end{subequations}
Since the QCD correction factors for the diagrams including internal
charm and up quarks are different from that for the top loop, we have
included the extra QCD factors $\hat{\eta}_{1,2}$ in $A_{\rm SM}(K)$
\cite{kkbarQCD}.
Explicit forms of $\eta_K$ and $\hat{\eta}_{1,2}$ are given in
\ref{sec:qcdfactors}.
The SUSY contributions $A_{\chi^\pm}(K)$ and $A_{\wt{g}}(K)$ are
obtained from the formulae for \bb mixing \ref{eq:bbcno} and
\ref{eq:bbgl} by an appropriate change of the flavor indices.
Then the \bb mass splitting $\Delta M_B$ and
the CP violation parameter $\epsilon_K$ are obtained from $M_{12}(B)$
and  $M_{12}(K)$ as
\begin{subequations}
\begin{eqnarray}
  \Delta M_B &=& 2 | M_{12}(B) | ~,
\\
  \epsilon_K &=& \e^{i\pi/4} \frac{ {\rm Im} M_{12}(K) }
                                  {\sqrt{2} \Delta M_K } ~,
\label{eq:epsilonK}
\end{eqnarray}
\end{subequations}
respectively.
The experimental value for the $K_L$--$K_S$ mass splitting is
given as $\Delta M_K = 3.51 \times 10^{-12}$ MeV, and we have used the
experimental result $\Delta \Gamma_K \approx -2\Delta M_K$ in
Eq.~\ref{eq:epsilonK}.
Note that the contributions from penguin diagrams is omitted to derive
\ref{eq:epsilonK}.
In the minimal SUGRA model, as well as SM, the $\epsilon_K$ is estimated
to be dominated by the box contributions.

\section{Numerical results}
\label{sec:results}

Following the method described in Ref.~\cite{gna}, we investigate the
three dimensional parameter space $\{ m_0,~ M_{gX}~, A_X \}$ within the
ranges $m_0, M_{gX} < 2$ TeV and $|A_X| < 5 m_0$ for a given set of
$\tan\beta$ and CKM parameters $|V_{us}|$, $|V_{cb}|$,
$|V_{ub}|/|V_{cb}|$ and $\delta_{13}$, where $\delta_{13}$ is the CP
violating phase in the standard parametrization \cite{PDG} and is
defined as $\e^{-i\delta_{13}} = V_{ub}/|V_{ub}|$.
Then we repeat the whole procedure varying $\tan\beta$ and the CKM
parameters.
The top quark mass is fixed to $m_t = 175$ GeV at the electroweak scale
\cite{top}.
In order to obtain the allowed region in the parameter space, we require
each calculated point to satisfy the following phenomenological
constraints \cite{PDG} beside the condition for the radiative
electroweak symmetry breaking scenario \cite{radbr}:
\begin{enumerate}
\item \bsg inclusive branching ratio.
  It is known that the \bsg branching ratio is approximately independent
  of $\delta_{13}$ and gives a unique constraint on the SUSY parameter
  space \cite{bbmr,bsgamma,go}.
  For the detail procedure to put a constraint on the SUSY parameter
  space, see Ref.~\cite{go}.
  The measurement by CLEO \cite{cleo} requires $1\times 10^{-4} <
  {\rm Br}( b \rightarrow s \gamma ) < 4.2\times 10^{-4}$;
\item The mass of any charged SUSY particle is larger than 45 GeV;
\item All sneutrino mass are larger than 41 GeV;
\item The gluino mass is larger than 100 GeV \cite{CDF};
\item Neutralino search results at LEP \cite{LEP}, which require
$\Gamma(Z \rightarrow \chi \chi)< 8.4$ MeV,
${\rm Br}(Z \rightarrow \chi \chi')$,
${\rm Br}(Z \rightarrow \chi' \chi')< 2 \times 10^{-5}$,
where $\chi$ is the lightest neutralino and
$\chi'$ is any neutralino other than the lightest one;
\item The lightest SUSY particle (LSP) is neutral;
\item The condition for not having a charge or color
  symmetry breaking vacuum \cite{aterm}.
\end{enumerate}
In the following, we show our results for $\theta_B$, $\Delta M_B$ and
$\epsilon_K$.

\subsection{$\theta_B$}

In \reffig{fig:reM12-imM12.tan03}, we show the complex value of
$M_{12}(B)$ for fixed $\tan\beta = 3$ and CKM parameters
$|V_{us}|=0.221$, $|V_{cb}|=0.041$ and $|V_{ub}|/|V_{cb}|=0.08$ with
four choices of $\delta_{13}=\pi/6$, $\pi/3$, $\pi/2$ and $2\pi/3$.
The axes are normalized to the prefactor $\hat{B}_B \eta_B f_B^2 M_B /
384\pi^2$ in \ref{eq:bbtotal}.
Each dot shows the value of $M_{12}(B)$ in the minimal SUGRA model and
each cross represents the SM value.
We see that all SUGRA points lie along the line connecting the origin
and the corresponding SM point, which shows that $\theta_B$ in the
minimal SUGRA model is equal to the SM value with the same CKM
parameters.
This fact is known previously \cite{bbbar,bcko} by analyses with
the approximate solutions of the RGEs where Yukawa couplings other than
that of top quark are neglected.
Our numerical result confirms and extends the previous analyses on this
point.
We have also checked that the result is independent of $\tan\beta$.
Phenomenologically, this has an important consequence that the CP
asymmetry measurements in various $B$ decay modes including
$B \rightarrow J/\psi\ K_S$ gives a direct information on the $\rho$ and
$\eta$ parameters just as in SM even if there are new contributions to
$\Delta M_B$ and/or $\epsilon_K$;
if the phase in the new contributions to $M_{12}(B)$ were different from
the phase of the SM contribution, the relation between the CP asymmetries
in $B$ decays and the CKM parameters would be modified and hence one
could not read CKM parameters directly from the measured CP asymmetries.

\subsection{$\Delta M_B$}

As can be seen in \reffig{fig:reM12-imM12.tan03}, SUSY and charged Higgs
contributions to the magnitude of $M_{12}(B)$ are all {\em constructive}
in the whole allowed parameter space.
We show the ratio of $\Delta M_B$ in the minimal SUGRA model
$\Delta M_B({\rm SUGRA})$ to the SM value $\Delta M_B({\rm SM})$ as
functions of the charged Higgs mass $m_{H^\pm}$ and the lighter
scalar top mass $m_{\wt{t}_1}$ for $\tan\beta=3$ and $10$ in
\reffig{fig:rxd.tan03} -- \reffig{fig:rxdstop.tan10}.
$\delta_{13}$ is fixed to $\pi/3$ and other CKM parameters are the
same as those in \reffig{fig:reM12-imM12.tan03}.
Each solid line in \reffig{fig:rxd.tan03} and \reffig{fig:rxd.tan10}
shows the value in THDM II with the same $\tan\beta$ and CKM parameters.
For a small $\tan\beta=3$, main non-SM contributions to $\Delta M_B$
come from  both the charged Higgs loop \ref{eq:bbhc} and the
chargino loop \ref{eq:bbcno} while the gluino contribution
\ref{eq:bbgl} is relatively small.
The total $\Delta M_B({\rm SUGRA})$ increases by $\sim$20\% of the SM
value for a charged Higgs mass $\lsim$ 300 GeV.
On the other hand, for a large $\tan\beta=10$, the charged Higgs
contribution is suppressed as $\sim 1/\tan^2\beta$.
In that case, the dominant non-SM contribution comes from the chargino
loop only.
\reffig{fig:rxdstop.tan03} and \reffig{fig:rxdstop.tan10}
show that a relatively light scalar top is necessary for a large
chargino -- scalar-top loop contribution for both choices of
$\tan\beta$.

In order to investigate the CKM parameter dependences of
$\Delta M_B({\rm SUGRA})/\Delta M_B({\rm SM})$, we varied the CKM
parameters within the range $|V_{ub}|/|V_{cb}|=0.08\pm 0.03$ and
$0<\delta_{13}<\pi$.
We find that the change of the non-SM part of $M_{12}(B)$ normalized to
the SM value is less than $O(10^{-4})$.
Combining with the result on $\theta_B$, one can see that the whole
$(\rho,~ \eta)$ dependence of the complex number $M_{12}(B)$ is common
to the SM and the non-SM parts and is canceled out in the ratio
$\Delta M_B({\rm SUGRA})/\Delta M_B({\rm SM})$ in a good accuracy.

The lower and the upper bounds of the ratio $\Delta M_B({\rm SUGRA}) /
\Delta M_B({\rm SM})$ in the parameter space $\{ m_{H^\pm},~
\tan\beta \}$ are shown in \reffig{fig:rxd.min.cont} and
\reffig{fig:rxd.max.cont}, respectively.
Here we fix the CKM parameters to the same values as those in
\reffig{fig:rxd.tan03} -- \reffig{fig:rxdstop.tan10} since the result
does not depend on the choice of the CKM parameters.
The range of the $\tan\beta$ we have scanned is $2\lsim \tan\beta \lsim
55$. For the values of $\tan\beta$ smaller or larger than this
range the Yukawa coupling constant for top or bottom/tau
blows up below the grand unification scale%
\footnote{
Precisely speaking, our calculation does not apply for very large
$\tan\beta \sim 55$ where the large bottom Yukawa coupling constant
induces the new operators involving the right-handed bottom quark.
}.
We see that the largest enhancement on $\Delta M_B$ ($\Delta M_B({\rm
  SUGRA})/\Delta M_B({\rm SM}) \gsim 1.2$) is realized for small
$\tan\beta \lsim 4$ and  $m_{H^\pm} \lsim 400$ GeV.
In this parameter region, a relatively light scalar top $m_{\wt{t}_1}
\lsim 200$ GeV also exists and hence both charged Higgs loop and
chargino and scalar top loop contribute to $\Delta M_B({\rm SUGRA})$
sizably.

Let us now consider what would change if LEP II should not find any
SUSY signal.
Since the upper bound of $\Delta M_B({\rm SUGRA})/\Delta M_B({\rm SM})$
for given $\tan\beta$ and $m_{H^\pm}$ is essentially determined by the
lower bounds for the masses of SUSY particles, chargino and/or scalar
top, in particular, a parameter region with relatively large
$\Delta M_B({\rm SUGRA})/\Delta M_B({\rm SM})$ is excluded if the lower
bound of the SUSY particle masses is raised to $\sim$90 GeV.
As a result, the upper bound of
$\Delta M_B({\rm SUGRA})/\Delta M_B({\rm SM})$ shown in
\reffig{fig:rxd.max.cont} decreases considerably,
while the lower bound of the charged Higgs mass for each
$\tan\beta$ is also raised.
On the other hand, the lower bound of
$\Delta M_B({\rm SUGRA})/\Delta M_B({\rm SM})$ shown in
\reffig{fig:rxd.min.cont} is insensitive to the lower bound of the SUSY
particle masses
because the bound is essentially determined by the requirement of the
radiative electroweak symmetry breaking.
Therefore, the only change in \reffig{fig:rxd.min.cont} after the LEP II
constraint is that the excluded region of the charged Higgs mass is
extended by $\sim$50 GeV (see Ref.~\cite{go}),
though the contours themselves are not changed much.

\subsection{$\epsilon_K$}

\reffig{fig:rek.tan03} -- \reffig{fig:rekstop.tan10} show scatter plots
of the ratio $|\epsilon_K({\rm SUGRA})| / |\epsilon_K({\rm SM})|$.
In comparison with the corresponding figures for
$\Delta M_B({\rm SUGRA}) / \Delta M_B({\rm SM})$,
\reffig{fig:rxd.tan03} -- \reffig{fig:rxdstop.tan10}, respectively, we
see that $|\epsilon_K({\rm SUGRA})| / |\epsilon_K({\rm SM})|$ and
$\Delta M_B({\rm SUGRA}) / \Delta M_B({\rm SM})$ have quite similar
characteristics.
We find actually that there is a linear relation between $\Delta
M_B({\rm SUGRA}) / \Delta M_B({\rm SM})$ and $|\epsilon_K({\rm SUGRA})|
/ |\epsilon_K({\rm SM})|$,
which is shown in \reffig{fig:rxd-rek.tan03}.
Here, CKM parameters are fixed to the same values as those in
\reffig{fig:rxd.tan03} with $\tan\beta = 3$.
This linear relation comes from the fact that the SUSY contributions to
both $M_{12}(B)$ and $M_{12}(K)$ are dominated by the box diagram with
the scalar top and the chargino loop, hence the enhancement factor is
common.
The small deviation of
$\frac{\Delta M_B({\rm SUGRA})/\Delta M_B({\rm SM})}%
{|\epsilon_K({\rm SUGRA})| / |\epsilon_K({\rm SM})|}$ from unity seen in
\reffig{fig:rxd-rek.tan03} is due to the contributions from charm quark.
In fact, we have checked that the enhancement factors for $\Delta M_B$
and $\epsilon_K$ coincide with each other if we neglect the charm quark
contributions to $\epsilon_K$ (see \ref{eq:kksm} and \ref{eq:kkhc}).
This fact is previously noticed in Ref.~\cite{bcko}, in which a
simplified form of squark mass matrices is assumed%
\footnote{
The ratio $\Delta M_B({\rm SUGRA}) / \Delta M_B({\rm SM})$ corresponds
to the parameter $R$ in Ref.~\cite{bcko}.
}.
We have numerically confirmed this point in the minimal SUGRA model.

\section{Conclusions}
\label{sec:conclusion}

In this paper we have made an extensive analysis on \bb mixing and
$\epsilon_K$ parameter in the minimal SUGRA model.
We have found that the present experimental constraints including the
recent measurement of \bsg branching ratio still allows for $+20$\%
deviation from the SM of both $\Delta M_B$ and $|\epsilon_K|$.
We also found that the enhancement factors
for $\Delta M_B$ and $|\epsilon_K|$ have a strong correlation.
We have seen that the dependence of the $\Delta M_B$ from the SUSY
contributions on the CKM matrix element is the same as that of the SM in
a very good accuracy so that the ratio does not depend on $\rho$ and
$\eta$.

Let us discuss implications of these results to constrain the SUSY
parameter space.
Since the effect of the new particles is at most 20$\sim$30\%, the
present constraints from $\Delta M_B$ and $\epsilon_K$ are not
very strong.
This is because the CKM parameters $(\rho,~\eta)$ are not determined
precisely from the other measurements.
Since we do not assume the SM, the only available information on
$(\rho,~\eta)$ is given by the measurement of $|V_{ub}|/|V_{cb}|$
which corresponds to $\sqrt{\rho^2 + \eta^2} = 0.36 \pm 0.14$.
The situation, however, will change when the CKM parameters are more
precisely determined from the measurements of the CP asymmetry in
$B$ decays at the future $B$-factories.
Since the phase of the \bb mixing amplitude in the minimal SUGRA is
the same as that in the SM, the CP asymmetries in $B$-decays such as
$B \rightarrow J/\psi\ K_S$ are directly related to the CKM
parameters just as in the SM case.
Therefore, it will be possible to extract 10$\sim$20\% effects from the
new particles after the CKM parameters are determined with enough
precision in the future.
It is thus important to measure CP violating
asymmetries in various modes of $B$ decay, not just in $B \rightarrow
J/\psi\ K_S$ mode, and to reduce the ambiguities on the hadron matrix
elements $f_B$, $\hat{B}_B$ and $\hat{B}_K$ from theoretical and/or
experimental improvements.
It is interesting to note that the parameter region in the $\tan\beta$
-- $m_{H^\pm}$ space which has the largest enhancements in $\Delta M_B$
and $\epsilon_K$ corresponds to relatively small values of $\tan\beta$
and $m_{H^\pm}$, \ie $\tan\beta \lsim 10$ and $m_{H^\pm} \lsim 300$ GeV.
%
%This parameter region is accessible in the future $B$-factory experiments
%though Higgs search might be difficult at the LHC experiments
%\cite{atlas} in this region if the LEP II experiments fail to find any
%signal of the Higgs boson.
%
This region roughly coincides with the parameter region in which the
Higgs search might be difficult at the LHC experiments \cite{atlas}
if the LEP II experiments fail to find any signal of the Higgs boson.
It may be possible, however, that the $B$-factory experiments will find
whether this parameter region is favored or not before the LHC
experiments start.

\subsection*{Acknowledgment}

The authors would like to thank K. Hikasa and J. Arafune for carefully
reading the manuscript and giving useful comments.
The work of Y.~O. is supported in part by the Grant-in-aid
for Scientific Research from the Ministry of Education,
Science and Culture of Japan.

\appendix

\section{Functions from loop integrals}
\label{sec:ilf}

The Inami-Lim functions used in Eqs.~\ref{eq:bbsm}--\ref{eq:bbgl}
and Eqs.~\ref{eq:kksm}, \ref{eq:kkhc} are the following:
\begin{eqnarray}
F_1(x) &=& \frac{1}{(x-1)^2}
\left\{
  \frac{3 x^3}{2(x-1)} \log x + x - \frac{11}{4} x^2 + \frac{1}{4} x^3
\right\} ~,
\\
F_2(x) &=&
  \log x - \frac{3 x}{4(x-1)} \left\{ \frac{x}{x-1} \log x - 1 \right\} ~,
\\
G_0(x,y) &=&
\frac{1}{x-y} \left\{ \frac{x}{(x-1)^2}\log x - \frac{1}{x-1}
                      - ( x \leftrightarrow y ) \right\} ~,
\\
G_1(x,y) &=&
\frac{1}{x-y} \left\{ \frac{x^2}{(x-1)^2}\log x - \frac{1}{x-1}
                      - ( x \leftrightarrow y ) \right\} ~,
\\
G'_0(x,y,z) &=&
\frac{1}{x-y}
 \left\{   \frac{1}{x-z}
           \left[   \frac{x}{(x-1)}\log x
                  - \frac{z}{(z-1)}\log z \right]
         - ( x \leftrightarrow y ) \right\} ~,
\\
G'_1(x,y,z) &=&
\frac{1}{x-y}
 \left\{   \frac{1}{x-z}
           \left[   \frac{x^2}{(x-1)}\log x
                  - \frac{z^2}{(z-1)}\log z \right]
         - ( x \leftrightarrow y ) \right\} ~.
\end{eqnarray}

\section{QCD factors}
\label{sec:qcdfactors}

We use the following formulae for the QCD factors $\eta_B$, $\eta_K$ and
$\hat{\eta}_{1,2}$ in \ref{eq:bbtotal}, \ref{eq:kktotal} and
\ref{eq:kksm}, respectively, which are obtained with one-loop calculations
\cite{kkbarQCD}.
We have neglected the threshold corrections near the electroweak scale:
\begin{eqnarray}
  \eta_B &=& \alpha_s(m_W)^{6/23} ~,
\label{eq:qcdeta}
\\
  \eta_K &=& \alpha_s(m_c)^{6/27}
         \left(\frac{\alpha_s(m_b)}{\alpha_s(m_c)}\right)^{6/25}
         \left(\frac{\alpha_s(m_W)}{\alpha_s(m_b)}\right)^{6/23} ~,
\label{eq:qcdetaK}
\\
  \hat{\eta}_1 &=&
  \frac{3}{2} \left(\frac{\alpha_s(m_b)}{\alpha_s(m_c)}\right)^{-18/25}
              \left(\frac{\alpha_s(m_W)}{\alpha_s(m_b)}\right)^{-18/23}
\nonumber\\&&
-             \left(\frac{\alpha_s(m_b)}{\alpha_s(m_c)}\right)^{-36/25}
              \left(\frac{\alpha_s(m_W)}{\alpha_s(m_b)}\right)^{-36/23}
\nonumber\\&&
+ \frac{1}{2} \left(\frac{\alpha_s(m_b)}{\alpha_s(m_c)}\right)^{-54/25}
              \left(\frac{\alpha_s(m_W)}{\alpha_s(m_b)}\right)^{-54/23} ~,
\\
  \hat{\eta}_2 &=&
  \frac{2\pi}{\alpha_s(m_W)\log{x_c}}
  \left\{
  \frac{ 9}{ 7} \left(\frac{\alpha_s(m_b)}{\alpha_s(m_c)}\right)^{7/25}
                \left(\frac{\alpha_s(m_W)}{\alpha_s(m_b)}\right)^{5/23}
\right.
\nonumber\\&&
+ \frac{35}{18} \left(\frac{\alpha_s(m_W)}{\alpha_s(m_b)}\right)^{5/23}
+ \frac{ 6}{11} \left(\frac{\alpha_s(m_b)}{\alpha_s(m_c)}\right)^{-11/25}
                \left(\frac{\alpha_s(m_W)}{\alpha_s(m_b)}\right)^{-13/23}
\nonumber\\&&
- \frac{12}{143}\left(\frac{\alpha_s(m_W)}{\alpha_s(m_b)}\right)^{-13/23}
- \frac{ 3}{29} \left(\frac{\alpha_s(m_b)}{\alpha_s(m_c)}\right)^{-29/25}
                \left(\frac{\alpha_s(m_W)}{\alpha_s(m_b)}\right)^{-31/23}
\nonumber\\&&
\left.
+ \frac{ 6}{899}\left(\frac{\alpha_s(m_W)}{\alpha_s(m_b)}\right)^{-31/23}
- \frac{4362}{2015}
  \right\} ~.
\end{eqnarray}

%%%%%%%%%%%%%%%%%%%%%%%%%%%%%%%%%%%%%%%%%%%%%%%%%%%%%%%%%%%%%%%%%%%%%%

%%%%%%%%%%%%%%%%%%%%%%%%%%%%%%%%%%%%%%%%%%%%%%%%%%%%%%%%%%%%%%%%%%%%%%

\newcommand{\inputfigure}[3]{%
\begin{figure}[hbtp]
  \begin{center}
    \leavevmode
    \makebox[0cm]{
    \epsfile{file=#1}
    }
  \end{center}
  \caption{#2}
  \label{#3}
\end{figure}
}

\newcommand{\CaptionReIm}{%
$M_{12}(B)$ normalized to $\hat{B}_B \eta_B f_B^2 M_B / 384\pi^2$ for
$\delta_{13}=\pi/6$, $\pi/3$, $\pi/2$ and $2\pi/3$ with fixed $m_t =
175$ GeV, $\tan\beta = 3$, $|V_{us}|=0.221$, $|V_{cb}|=0.041$ and
$|V_{ub}|/|V_{cb}|=0.08$.
The cross represents the standard model value.
}
\newcommand{\CaptionRxdTanIII}{%
Ratio of $\Delta M_B$ in the minimal SUGRA model to the standard model
value as a function of the charged Higgs mass with fixed $m_t = 175$
GeV, $\tan\beta = 3$, $|V_{us}|=0.221$, $|V_{cb}|=0.041$,
$|V_{ub}|/|V_{cb}|=0.08$ and $\delta_{13} = \pi/3$.
The solid line shows the value in THDM II.
}
\newcommand{\CaptionRxdStopTanIII}{%
Ratio of $\Delta M_B$ in the minimal SUGRA model to the standard model
value as a function of the lighter scalar top mass with the same
parameters as those in \reffig{fig:rxd.tan03}.
}
\newcommand{\CaptionRxdTanX}{%
Same as \reffig{fig:rxd.tan03} for $\tan\beta = 10$.
}
\newcommand{\CaptionRxdStopTanX}{%
Same as \reffig{fig:rxdstop.tan03} for $\tan\beta = 10$.
}
\newcommand{\CaptionMinCont}{%
Contour plot for the minimal value of $\Delta M_B({\rm SUGRA}) / \Delta
M_B({\rm SM})$
on $m_{H^\pm}$--$\tan\beta$ plane.
Each number attached to each contour line represents the value of
$\Delta M_B({\rm SUGRA}) / \Delta M_B({\rm SM})$.
}
\newcommand{\CaptionMaxCont}{%
Contour plot for the maximal value of $\Delta M_B({\rm SUGRA}) / \Delta
M_B({\rm SM})$
on $m_{H^\pm}$--$\tan\beta$ plane.
Each number attached to each contour line represents the value of
$\Delta M_B({\rm SUGRA}) / \Delta M_B({\rm SM})$.
}
\newcommand{\CaptionRekTanIII}{%
Ratio of $|\epsilon_K|$ in the minimal SUGRA model to the standard model
value as a function of the charged Higgs mass with the same
parameters as those in \reffig{fig:rxd.tan03}.
}
\newcommand{\CaptionRekStopTanIII}{%
Ratio of $|\epsilon_K|$ in the minimal SUGRA model to the standard model
value as a function of the lighter scalar top mass with the same
parameters as those in \reffig{fig:rxd.tan03}.
}
\newcommand{\CaptionRekTanX}{%
Same as \reffig{fig:rek.tan03} for $\tan\beta = 10$.
}
\newcommand{\CaptionRekStopTanX}{%
Same as \reffig{fig:rekstop.tan03} for $\tan\beta = 10$.
}
\newcommand{\CaptionRxdRekTanIII}{%
Correlation between the enhancement factors
$\Delta M_B({\rm SUGRA}) / \Delta M_B({\rm SM})$ and
$|\epsilon_K({\rm SUGRA})| / |\epsilon_K({\rm SM})|$.
Parameters are fixed to the same values as those in
\reffig{fig:rxd.tan03}.
}

\newpage
\section*{Figure Captions}
\typeout{***** Figure Captions *****}

\begin{list}{\bf FIG.~??}{\relax}
\Fig{fig:reM12-imM12.tan03} \CaptionReIm
\Fig{fig:rxd.tan03}         \CaptionRxdTanIII
\Fig{fig:rxdstop.tan03}     \CaptionRxdStopTanIII
\Fig{fig:rxd.tan10}         \CaptionRxdTanX
\Fig{fig:rxdstop.tan10}     \CaptionRxdStopTanX
\Fig{fig:rxd.min.cont}      \CaptionMinCont
\Fig{fig:rxd.max.cont}      \CaptionMaxCont
\Fig{fig:rek.tan03}         \CaptionRekTanIII
\Fig{fig:rekstop.tan03}     \CaptionRekStopTanIII
\Fig{fig:rek.tan10}         \CaptionRekTanX
\Fig{fig:rekstop.tan10}     \CaptionRekStopTanX
\Fig{fig:rxd-rek.tan03}     \CaptionRxdRekTanIII
\end{list}

\newpage
\section*{Figures}
\typeout{***** Figures *****}

\inputfigure{reM12-imM12.tan03.eps,height=15cm}{\CaptionReIm}{fig:reM12-imM12.tan03}
\begin{subfigures}
\inputfigure{rxd.tan03.eps,height=15cm}{\CaptionRxdTanIII}{fig:rxd.tan03}
\inputfigure{rxdstop.tan03.eps,height=15cm}{\CaptionRxdStopTanIII}{fig:rxdstop.tan03}
\inputfigure{rxd.tan10.eps,height=15cm}{\CaptionRxdTanX}{fig:rxd.tan10}
\inputfigure{rxdstop.tan10.eps,height=15cm}{\CaptionRxdStopTanX}{fig:rxdstop.tan10}
\end{subfigures}
\begin{subfigures}
\inputfigure{rxd.min.cont.eps,height=15cm}{\CaptionMinCont}{fig:rxd.min.cont}
\inputfigure{rxd.max.cont.eps,height=15cm}{\CaptionMaxCont}{fig:rxd.max.cont}
\end{subfigures}
\begin{subfigures}
\inputfigure{rek.tan03.eps,height=15cm}{\CaptionRekTanIII}{fig:rek.tan03}
\inputfigure{rekstop.tan03.eps,height=15cm}{\CaptionRekStopTanIII}{fig:rekstop.tan03}
\inputfigure{rek.tan10.eps,height=15cm}{\CaptionRekTanX}{fig:rek.tan10}
\inputfigure{rekstop.tan10.eps,height=15cm}{\CaptionRekStopTanX}{fig:rekstop.tan10}
\end{subfigures}
\inputfigure{rxd-rek.tan03.eps,height=15cm}{\CaptionRxdRekTanIII}{fig:rxd-rek.tan03}

%%%%%%%%%%%%%%%%%%%%%%%%%%%%%%%%%%%%%%%%%%%%%%%%%%%%%%%%%%%%%%%%%%%%%%

\end{document}